\documentclass[preprint,aps,draft,showpacs]{revtex4}

\usepackage{graphicx}
\usepackage{dcolumn}
\usepackage{bm}

\begin{document}

\title{Remarks on Schwinger Pair Production
by Charged Black Holes}

\author{Sang Pyo Kim}\email{sangkim@kunsan.ac.kr}

\affiliation{Department of Physics, Kunsan National University,
Kunsan 573-701, Korea}

\affiliation{Asia Pacific Center for Theoretical Physics, Pohang
790-784, Korea}

\author{Don N. Page}\email{don@phys.ualberta.ca}

\address{Theoretical Physics Institute, Department of Physics, University
of Alberta, Edmonton, Alberta, Canada T6G 2J1}

\date{\today}
\begin{abstract}
We introduce a canonical method for pair production by
electromagnetic fields. The canonical method in the
space-dependent gauge provides pair-production rate even for
inhomogeneous fields. Further, the instanton action including all
corrections leads to an accurate formula for the pair-production
rate. We discuss various aspects of the canonical method and
clarify terminology for pair production. We study pair production
by charged black holes first by finding states of the field
equation that describe pair production and then by applying the
canonical method.
\end{abstract}

\maketitle

\section{Introduction}

Strong quantum electrodynamics (QED) significantly differs from
weak perturbative QED. Pairs of bosons and fermions are produced
from external electric fields exceeding the critical strength $E_c
= m^2c^3/\hbar q$. Vacuum fluctuations polarize the vacuum and the
instability therefrom leads to pairs of particles
\cite{heisenberg,schwinger}.  This pair production of charged
particles and antiparticles is a nonpertubative effect of QED. The
recent development of X-ray free electron laser technology may
make a test possible for pair production and attracts interests in
this strong QED \cite{ringwald,roberts}. Other natural phenomena
associated with strong electromgnetic fields are astrophysical
objects. For instance, pulsars have magnetic fields comparable
with or above the critical strength, and the magnetic fields of
magnetars exceed the critical strength by several order of
magnitude. Likewise, charged black holes, especially the extremal
ones, can have electric fields far greater than the critical
strength. Recently, Ruffini {\it et al.}
\cite{ruffini1,ruffini2,ruffini3} proposed a mechanism for gamma
rays burst through pair production and annihilation from charged
black holes.

In this paper we introduce the recent canonical method for pair
production \cite{kim-page1,kim-page2}, which can be applied to
inhomogeneous electromagnetic fields such as charged black holes.
In fact, the electric fields by charged black holes are
inhomogeneous in that the direction is radial and the field
strength depends on the radial distance. The pair-production rate
by the proper time method was derived for static uniform fields
\cite{heisenberg,schwinger}. However, this formula was used to
calculate the pair-production rate by charged back holes
\cite{ruffini1,ruffini2,ruffini3,gibbons}. Thus a field
theoretical method is needed to apply even to such inhomogeneous
fields. Our canonical method is an extension of the early idea
that particle production is related with tunneling processes in
the space-dependent gauge
\cite{nikishov,brezin,popov,casher,brout}. In the space-dependent
gauge of a static uniform electric field, each mode of the
Klein-Gordon or Dirac or vector field equations has a potential
barrier. It is the observation that the tunneling probability has
relation with pair production and the no-tunneling probability is
related with the vacuum persistence. Further, the tunneling
probability is determined by the instanton action including all
quantum corrections. Thus this canonical method can readily be
applied to static inhomogeneous electric fields, in particular,
charged black holes.

The organization of this paper is as follows. In Sec. II we review
the vacuum polarization and pair production in proper time method.
In Sec. III we use the time- and space-dependent gauges to obtain
the pair-production rate by a uniform electric field. We compare
the similarity and differences between two gauges. It is shown
that particle production is related with the tunneling probability
and the vacuum persistence is determined by the no-tunneling
probability. In Sec. IV we discuss the spin effect on pair
production and clarify the terminology of pair-production
probability, pair-production rate and the mean number of pairs. In
particular, the tunneling probability is given by the instanton
action including all quantum corrections. In Sec. V we study the
Klein-Gordon equation first in Minkowski spacetime and then in
charged black hole background. It is shown that there are
quasi-stationary states describing purely outgoing waves with
complex frequencies corresponding to pair-production states.
Finally, we apply the canonical method to calculate the
pair-production rates by charged black holes.

\section{Vacuum Polarization and Pair Production}

At the dawning of quantum field theory, Heisenberg and Euler
calculated perturbatively the effective action for uniform
electromagnetic fields \cite{heisenberg}. Later, Schwinger found
the one-loop effective action for fermions of spin $1/2$ in
uniform electromagnetic fields by using the proper time method
\cite{schwinger}:
\begin{equation}
{\cal L}_{e} = - {\cal F} - \frac{1}{8 \pi^2} \int_0^{\infty} ds
\frac{e^{-m^2s}}{s^3} \Biggl[(es)^2 {\cal G} \frac{{\rm Re} \cosh
(esX) }{{\rm Im} \cosh (esX)} - 1 - \frac{2}{3} (es)^2 {\cal F}
\Biggr], \label{one-loop}
\end{equation}
where
\begin{equation}
{\cal F} = \frac{1}{4} F_{\mu \nu} F^{\mu \nu} = \frac{1}{2} ({\bf
B}^2 - {\bf E}^2),  \quad {\cal G} = \frac{1}{4} F^*_{\mu \nu}
F^{\mu \nu} = {\bf E} \cdot {\bf B},
\end{equation}
and
\begin{equation}
X = [2 ({\cal F} + i {\cal G})]^{1/2} = X_r + i X_i.
\end{equation}
The one-loop effective action can be written in the form
\begin{equation}
{\cal L}_{e} = - {\cal F} - \frac{1}{8 \pi^2} \int_0^{\infty} ds
\frac{e^{-m^2s}}{s^3} \Biggl[(es)^2 {\cal G} \coth (esX_r) \cot
(esX_i) - 1 - \frac{2}{3} (es)^2 {\cal F} \Biggr],
\end{equation}
which has simple poles at $s_n = (n \pi)/(eX_i)$, $(n = 1, 2,
\cdots)$. These poles contribute the imaginary part
\begin{equation}
2 {\rm Im} {\cal L}_{e} = \frac{1}{4 \pi} \sum_{n = 1}^{\infty}
\frac{e {\cal G}}{X_i s_n} e^{- m^2 s_n} \coth
\Bigl(\frac{X_r}{X_i} n \pi \Bigr). \label{form 1}
\end{equation}
The real part of the one-loop effective action gives rise to the
vacuum polarization
\begin{eqnarray}
{\rm Re}  {\cal L}_{e} &=& - {\cal F} - \frac{1}{8 \pi^2}
\int_0^{\infty} ds \frac{e^{-m^2s}}{s^3} \Biggl[ (es)^2 X_r X_i
\Bigl\{ \frac{1}{esX_i} - \sum_{k = 1}^{\infty} \frac{2^{2k} \vert
B_{2k} \vert }{(2k)!} \nonumber\\ &&\times (es X_i)^{2k-1} \coth
(esX_r) \Bigr\}- 1 -
\frac{2}{3} (es)^2 {\cal F} \Biggr] \nonumber\\
& = & - {\cal F} + \frac{2}{45} \Bigl( \frac{e^2}{4 \pi \hbar c
}\Bigr)^2 \frac{(\hbar/mc)^3}{mc^2} \Bigl[({\bf B}^2 - {\bf
E}^2)^2  + 7 ({\bf E} \cdot {\bf B})^2 \Bigr] + \cdots
\label{real}
\end{eqnarray}

For a pure electric field $({\bf B} = 0, X_r = 0, X_i = E)$, Eq.
(\ref{form 1}) leads to the pair-production rate
\begin{equation}
2 {\rm Im} {\cal L}_{e} = \frac{1}{4 \pi^3} \sum_{n = 1}^{\infty}
\Bigl(\frac{eE}{n}\Bigr)^2 e^{- \frac{n \pi m^2}{eE}}. \label{form
2}
\end{equation}
In the case of bosons, the pair-production rate becomes
\begin{equation}
2 {\rm Im} {\cal L}_{e}^b = \frac{1}{8 \pi^3} \sum_{n =
1}^{\infty} (-1)^{n +1} \Bigl(\frac{eE}{n}\Bigr)^2 e^{- \frac{n
\pi m^2}{eE}}. \label{form 3}
\end{equation}
These imaginary parts determine the vacuum persistence
(vacuum-to-vacuum transition)
\begin{equation}
\vert \langle 0, {\rm out} \vert 0, {\rm in} \rangle \vert^2 =
e^{- 2 VT {\rm Im} {\cal L}_{e}},
\end{equation}
and thus lead to pair production of particle and antiparticle. It
should be remarked that the pair-production rates  (\ref{form 1}),
(\ref{form 2}), (\ref{form 3}) and the vacuum polarization
(\ref{real}) are valid only for uniform fields. These formulae may
be used as long as fields vary slowly in region of interest. There
are many situations, such as charged black holes, where the fields
are strong but not uniform. In the next sections, using a
canonical method, we shall derive the pair-production rates valid
even for inhomogeneous fields.

\section{Canonical Method for Pair Production}

Pair production of charged particles and antiparticles can also be
understood in canonical theory. An advantage of the canonical
method is that it has a direct meaning of pair production compared
with the path integral methods. In the canonical method one
directly solves either the Klein-Gordon or Dirac equation. An
interesting point is that one can easily extend the result for
bosons to fermions by taking into account the Pauli blocking
effect due to spins. The Klein-Gordon equation for bosons is given
by
\begin{equation}
\Bigl[- \eta^{\mu \nu} \Bigl(\frac{\partial}{\partial x^{\mu}}
\Bigr) \Bigl(\frac{\partial}{\partial x^{\nu}} \Bigr) + m^2 \Bigr]
\Phi (t, {\bf x}) = 0.
\end{equation}
In this section we focus on the uniform electric field, which is
determined by either vector potential or scalar potential
\begin{equation}
{\bf E} = - \frac{\partial}{\partial t} {\bf A} - \nabla A_0.
\end{equation}
A uniform electric field has a time-dependent vector potential
\begin{equation}
{\bf A} = - E t {\bf k} \label{vec gauge}
\end{equation}
or space-dependent scalar potential
\begin{equation}
A_0 = - E z. \label{sca gauge}
\end{equation}

\subsection{Time-dependent Gauge}

In the time-dependent gauge (\ref{vec gauge}) with the electric
field in the $z$-direction, the Klein-Gordon equation for spin-0
bosons now takes the form
\begin{equation}
\Bigl[\partial_t^2 - \partial_{\perp}^2 - (\partial_z + i qEt)^2 +
m^2 \Bigr] \Phi = 0.
\end{equation}
The field is decomposed into Fourier mode, $\Phi = e^{i {\bf k}
\cdot {\bf x}} \phi_{\bf k}$ and results in a one-dimensional
equation in the time-direction
\begin{equation}
\Bigl[- \partial_t^2 - (k_z + qEt)^2 - (m^2 + {\bf k}_{\perp}^2)
\Bigr] \phi_{\bf k} = 0. \label{time eq}
\end{equation}
The above equation describes a particle moving over an upside down
oscillator potential with positive energy.  This equation is an
analog of fields in a time-dependent spacetime background. We may
apply the canonical method by Parker and DeWitt to calculate the
amount of particle creation \cite{parker1,parker2,dewitt}. The
idea is that an incoming positive frequency solution at the past
infinity scatters by the potential and separates into an outgoing
positive and negative frequency solutions at the future infinity.
The negative frequency solution leads to particle creation.
Indeed, Eq. (\ref{time eq}) has an incoming positive frequency
solution at $t \rightarrow - \infty$
\begin{equation}
\phi_{{\bf k}, in} (t) = \frac{1}{(8 qE)^{1/4}} \Bigl[i
\sqrt{\kappa} W(- a_{{\bf k}_{\perp}}, \tau) +
\frac{1}{\sqrt{\kappa}} W(- a_{{\bf k}_{\perp}}, - \tau) \Bigr],
\label{incoming}
\end{equation}
where
\begin{eqnarray}
a_{{\bf k}_{\perp}} = \frac{{\bf k}_{\perp}^2 + m^2}{2 qE}, \quad
 \tau = \sqrt{2qE} t, \quad
\kappa = \sqrt{1+ e^{- 2 \pi a_{{\bf k}_{\perp}} }} - e^{- \pi
a_{{\bf k}_{\perp}} }.
\end{eqnarray}
The incoming wave can be written in another form
\begin{equation}
\phi_{{\bf k}, in} (t) = \frac{1}{(8 qE)^{1/4}} \Bigl[ \frac{i}{2}
\Bigl( \kappa - \frac{1}{\kappa} \Bigr) E (- a_{{\bf k}_{\perp}},
\tau) + \frac{i}{2} \Bigl( \kappa + \frac{1}{\kappa} \Bigr) E^* (-
a_{{\bf k}_{\perp}}, \tau)  \Bigr].
\end{equation}
In fact, the incoming wave is a superposition of the outgoing
positive and negative frequency solution
\begin{equation}
\phi_{{\bf k}, in} (t) =  \mu_{\bf k} \phi_{{\bf k}, out} (t) +
\nu_{\bf k} \phi^*_{{\bf k}, out}, \label{td qm}
\end{equation}
where $\phi_{{\bf k}, out}$ is the outgoing positive frequency
solution given by
\begin{equation}
\phi_{{\bf k}, out} (t) = \frac{1}{(8 qE)^{1/4}} E^* (- a_{{\bf
k}_{\perp}}, \tau),
\end{equation}
and
\begin{equation}
\mu_{\bf k} = \frac{i}{2} \Bigl( \kappa + \frac{1}{\kappa} \Bigr),
\quad \nu_{\bf k} = \frac{i}{2} \Bigl( \kappa - \frac{1}{\kappa}
\Bigr).
\end{equation}
Equation (\ref{td qm}) is equivalent to the Bogoliubov
transformation between the operators in two asymptotic regions
\begin{equation}
\hat{b}_{{\bf k}, out} = \mu_{\bf k} \hat{b}_{{\bf k}, in} +
\nu_{\bf k}^* \hat{b}^{\dagger}_{{\bf k}, out},
\end{equation}

The number of created particles in pairs in the future infinity is
given by
\begin{equation}
\langle 0, {\rm in} \vert \sum_{\bf k} \hat{b}^{\dagger}_{{\bf k},
out} \hat{b}_{{\bf k}, out} \vert 0, {\rm in} \rangle = \sum_{{\bf
k}_{\perp}} \vert \nu_{\bf k}\vert^2 = \sum_{{\bf k}_{\perp}} e^{-
2 \pi a_{{\bf k}_{\perp}}}.
\end{equation}
Here the mean number of created pairs in the transverse mode ${\bf
k}_{\perp}$ is ${\cal N}_{{\bf k}_{\perp}} = \vert \nu_{\bf k}
\vert^2 \leq 1$. Then the vacuum persistence or vacuum-to-vacuum
transition is given by
\begin{equation}
\vert \langle 0, {\rm out} \vert 0, {\rm in} \rangle \vert^2 =
\prod_{{\bf k}_{\perp}} \frac{1}{\vert \mu_{\bf k}\vert^2},
\end{equation}
where
\begin{equation}
\mu_{\bf k} = ( 1+ e^{- 2 \pi a_{{\bf k}_{\perp}}})^{1/2}.
\end{equation}
Thus the vacuum persistence is related with the imaginary part of
the effective action as
\begin{eqnarray}
\vert \langle 0, {\rm out} \vert 0, {\rm in} \rangle \vert^2 =
\exp \Bigl[- \sum_{{\bf k}_{\perp}} \ln (1+ e^{- 2 \pi a_{{\bf
k}_{\perp}}}) \Bigr]  = \exp \Bigl[- 2 VT {\rm Im} {\cal L}_{e}^b
\Bigr]
\end{eqnarray}
Now the imaginary part of the effective action for bosons per
volume per time is given by
\begin{eqnarray}
2 {\rm Im} {\cal L}_{e}^b = \frac{1}{VT} \sum_{{\bf k}_{\perp}}
\ln (1+ e^{- 2 \pi a_{{\bf k}_{\perp}}}) = \frac{1}{(2 \pi)^d}
\sum_{n = 1}^{\infty} (-1)^{n +1}
\Bigl(\frac{eE}{n}\Bigr)^{(d+1)/2} e^{- \frac{n \pi m^2}{eE}}.
\end{eqnarray}
We may interpret Eq. (\ref{td qm}) quantum mechanically. In the
space with $x = - t$, $\phi_{{\bf k}, out}$ corresponds to an
incident wave and $\phi_{{\bf k}, out}^*$ to a reflected one. Thus
the transmission coefficient gives the vacuum persistence
\begin{eqnarray}
\Bigl| \frac{1}{\mu_{\bf k}} \Bigr|^2 &=& 1 - \Bigl|
\frac{\nu_{\bf k}}{\mu_{\bf k}} \Bigr|^2
 = \frac{1}{1 + e^{ - 2 \pi a_{{\bf k}_{\perp}}}}
= \vert \langle 0_{{\bf k}_{\perp}}, {\rm out} \vert 0_{{\bf
k}_{\perp}}, {\rm in} \rangle \vert^2
\end{eqnarray}
and the mean number of created pairs is given by the reflection
coefficient
\begin{equation}
| \nu_{\bf k}|^2 = e^{ - 2 \pi a_{{\bf k}_{\perp}}} = {\cal N
}_{{\bf k}_{\perp}}.
\end{equation}

\subsection{Space-dependent Gauge}

In the space-dependent gauge, i.e. the Coulomb gauge (\ref{sca
gauge}), the Klein-Gordon equation, after the mode-decomposition
$\Phi = e^{ i ({\bf k}_{\perp} \cdot {\bf x}_{\perp} - \omega t)}
\phi_{\omega {\bf k}_{\perp}}$, separates into the mode equation
\begin{equation}
\Bigl[- \partial_z^2 - (\omega + qEz)^2 + (m^2 + {\bf
k}_{\perp}^2) \Bigr] \phi_{\omega {\bf k}_{\perp}} = 0.
\label{space eq}
\end{equation}
This equation describes a particle moving under the potential
barrier in contrast with the over-barrier in the time-dependent
gauge. The tunneling wave function is given in terms of the
complex cylindrical function
\begin{equation}
\phi_{\omega {\bf k}_{\perp}} (\xi) = c E(a_{{\bf k}_{\perp}} ,
\xi),
\end{equation}
where
\begin{equation}
a_{{\bf k}_{\perp}} = \frac{{\bf k}_{\perp}^2 + m^2}{2 q E}, \quad
\xi = \sqrt{\frac{2}{qE}} (\omega + qE z).
\end{equation}
This wave function has the proper asymptotic forms at $z
\rightarrow - \infty$
\begin{equation}
\phi_{\omega {\bf k}_{\perp}} (\xi \rightarrow - \infty) = A_{{\bf
k}_{\perp}} \sqrt{\frac{2}{|\xi|}} e^{- \frac{i}{4} \xi^2} +
B_{{\bf k}_{\perp}} \sqrt{\frac{2}{|\xi|}} e^{+ \frac{i}{4}
\xi^2},
\end{equation}
and at $z \rightarrow + \infty$
\begin{equation}
\phi_{\omega {\bf k}_{\perp}} (\xi \rightarrow  \infty) = C_{{\bf
k}_{\perp}} \sqrt{\frac{2}{|\xi|}} e^{+ \frac{i}{4} \xi^2}.
\end{equation}
Here the coefficients are
\begin{equation}
A_{{\bf k}_{\perp}} = i c \sqrt{1 + e^{ 2 \pi a_{{\bf
k}_{\perp}}}}, \quad B_{{\bf k}_{\perp}} = ic e^{ \pi a_{{\bf
k}_{\perp}}}, \quad \quad C_{{\bf k}_{\perp}} = c.
\end{equation}
Then the tunneling probability is
\begin{equation}
P_{{\bf k}_{\perp}}^b = \Bigl|\frac{C_{{\bf k}_{\perp}}}{A_{{\bf
k}_{\perp}}} \Bigr|^2 = \frac{1}{ 1 + e^{ 2 \pi a_{{\bf
k}_{\perp}}}}, \label{tun pr}
\end{equation}
and the probability for no-tunneling
\begin{equation}
P_{{\bf k}_{\perp}}^{nb} = \Bigl|\frac{B_{{\bf
k}_{\perp}}}{A_{{\bf k}_{\perp}}} \Bigr|^2 = \frac{1}{1 + e^{ - 2
\pi a_{{\bf k}_{\perp}}}}. \label{notun pr}
\end{equation}
Note that the instanton action for tunneling
\begin{equation}
S_{{\bf k}_{\perp}} = \frac{1}{2} \oint dz \sqrt{ (m^2 + {\bf
k}_{\perp}^2)- (\omega + qEz)^2} = \pi a_{{\bf k}_{\perp}},
\end{equation}
after taking into account multi-instantons and
multi-anti-instantons contributions, leads to the total tunneling
probability
\begin{equation}
P_{{\bf k}_{\perp}}^b = \sum_{n = 1}^{\infty} (-1)^{n+1} e^{ - 2n
S_{{\bf k}_{\perp}}} = \frac{1}{1 + e^{2 S_{{\bf k}_{\perp}}}},
\end{equation}
and the total probability for no-tunneling
\begin{equation}
P_{{\bf k}_{\perp}}^{nb} = \sum_{n = 0}^{\infty} (-1)^{n} e^{ - 2n
S_{{\bf k}_{\perp}}} = \frac{1}{1 + e^{- 2 S_{{\bf k}_{\perp}}}}.
\end{equation}

The idea of Refs.
\cite{kim-page1,kim-page2,nikishov,brezin,popov,casher,brout} is
that the tunneling probability is related with pair production and
the no-tunneling probability with the vacuum-to-vacuum transition.
As in the case of the time-dependent gauge, the vacuum persistence
is given by
\begin{equation}
\vert \langle 0, {\rm out} \vert 0, {\rm in} \rangle \vert^2 =
\prod_{{\bf k}_{\perp}} P_{{\bf k}_{\perp}}^{nb} = \exp \Bigl[- 2
VT {\rm Im} {\cal L}_{e}^b \Bigr].
\end{equation}
Finally, the imaginary part of the effective action for bosons per
volume per time is given by
\begin{eqnarray}
2 {\rm Im} {\cal L}_{e}^b = \frac{1}{VT} \sum_{{\bf k}_{\perp}}
\ln (1+ e^{- 2 S_{{\bf k}_{\perp}}}) = \frac{1}{(2 \pi)^d} \sum_{n
= 1}^{\infty} (-1)^{n +1} \Bigl(\frac{eE}{n}\Bigr)^{(d+1)/2} e^{-
\frac{n \pi m^2}{eE}}.
\end{eqnarray}
In the fermion case, fermion pairs in the same state are
prohibited by the Pauli exclusion principle, so the tunneling
probability is limited to
\begin{equation}
P_{{\bf k}_{\perp}}^{f} = e^{ - 2 \pi a_{{\bf k}_{\perp}}},
\end{equation}
and the no-tunneling probability to
\begin{equation} P_{{\bf k}_{\perp}}^{nf} = 1 - e^{ - 2 \pi
a_{{\bf k}_{\perp}}}.
\end{equation}
Thus the imaginary part of the effective action for fermions per
volume per time is given by
\begin{eqnarray}
2 {\rm Im} {\cal L}_{e}^f = - \frac{2}{VT} \sum_{{\bf k}_{\perp}}
\ln (1 -  e^{- 2 S_{{\bf k}_{\perp}}}) = \frac{2}{(2 \pi)^d}
\sum_{n = 1}^{\infty} \Bigl(\frac{eE}{n}\Bigr)^{(d+1)/2} e^{-
\frac{n \pi m^2}{eE}}.
\end{eqnarray}

\section{Some Remarks on Canonical Method}

In this section we discuss some issues of pair production in the
canonical method. In Sec. III we derived the pair-production rate of
spin-0 bosons and spin-1/2 fermions. Here we introduce a recent
result extending the canonical method to spin particles
\cite{kim-page2}. We also clarify the different terminology for
particle production in literature. Finally, we show that the
canonical method based on the instanton action can readily be
generalized to inhomogeneous fields.

\subsection{Spin and Scattering Processes}

To treat the bosons with spin $\sigma \geq 1$ and fermions with
$\sigma \geq 3/2$, one has to solve the Klein-Gordon or Dirac
equation coupled with the vector field. Then the Klein-Gordon
equation for spin $\sigma$-particles takes the form
\begin{equation}
\Bigl[- \eta^{\mu \nu} \Bigl(\frac{\partial}{\partial x^{\mu}} + i
qA_{\mu} \Bigr) \Bigl(\frac{\partial}{\partial x^{\nu}} + i
qA_{\nu} \Bigr) + m^2  + 2 i \sigma qE \Bigr] \Phi (t, {\bf x}) =
0.
\end{equation}
After mode-decomposition, it reduces to
\begin{equation}
\Bigl[- \partial_z^2 - (\omega + qEz)^2 + (m^2 + {\bf
k}_{\perp}^2) + 2 i \sigma qE \Bigr] \phi_{\sigma \omega {\bf
k}_{\perp}} = 0.
\end{equation}
The tunneling wave function is now given by
\begin{equation}
\phi_{\omega {\bf k}_{\perp}} (\xi) = c E(a_{{\bf k}_{\perp}} ,
\xi),
\end{equation}
where
\begin{equation}
a_{{\bf k}_{\perp}} = \frac{{\bf k}_{\perp}^2 + m^2 + 2 i \sigma
qE}{2 q E}, \quad \xi = \sqrt{\frac{2}{qE}} (\omega + qE z).
\end{equation}
Denoting the asymptotic form of the wave function
\begin{equation}
\varphi_{\omega {\bf k}_{\perp}} (\xi) =  \sqrt{\frac{2}{|\xi|}}
e^{- \frac{i}{4} \xi^2},
\end{equation}
we find the wave function at $\xi = - \infty$
\begin{equation}
\phi_{\omega {\bf k}_{\perp}} = A_{{\bf k}_{\perp}}
\varphi_{\omega {\bf k}_{\perp}} - B_{{\bf k}_{\perp}}
\varphi_{\omega {\bf k}_{\perp}}^*,
\end{equation}
and at $\xi = + \infty$
\begin{equation}
\phi_{\omega {\bf k}_{\perp}} = C_{{\bf k}_{\perp}}
\varphi_{\omega {\bf k}_{\perp}}^*.
\end{equation}
In terms of group velocity, the flux conservation reads for bosons
\begin{equation}
\vert A_{{\bf k}_{\perp}} \vert^2 = \vert B_{{\bf k}_{\perp}}
\vert^2 + \vert C_{{\bf k}_{\perp}} \vert^2,
\end{equation}
and for fermions
\begin{equation}
\vert A_{{\bf k}_{\perp}} \vert^2 + \vert C_{{\bf k}_{\perp}}
\vert^2 = \vert B_{{\bf k}_{\perp}} \vert^2.
\end{equation}
Therefore, the reflection coefficient for bosons is
\begin{equation}
\Bigl| \frac{B_{{\bf k}_{\perp}}}{A_{{\bf k}_{\perp}}} \Bigr|^2 =
1 - \Bigl| \frac{C_{{\bf k}_{\perp}}}{A_{{\bf k}_{\perp}}}
\Bigr|^2,
\end{equation}
and for fermions
\begin{equation}
\Bigl| \frac{A_{{\bf k}_{\perp}}}{B_{{\bf k}_{\perp}}} \Bigr|^2 =
1 - \Bigl| \frac{C_{{\bf k}_{\perp}}}{B_{{\bf k}_{\perp}}}
\Bigr|^2.
\end{equation}
This result makes the difference of pair production between bosons
and fermions.

\subsection{Pair Production Rate and Mean Number}

Different terminology for pair production has been used in
literature. In some cases one has to take care of the different
quantities. Let us denote
\begin{equation}
w = 2 {\rm Im} {\cal L}_e
\end{equation}
per volume per time, then in the case of both electric and
magnetic fields this quantity becomes
\begin{eqnarray}
w_b &=& \frac{(qE)(qB)}{2 (2 \pi)^2} \sum_{n = 1}^{\infty}
(-1)^{n+1} \frac{1}{n} {\rm cosech} \Bigl(\frac{n \pi B}{E} \Bigr)
e^{- \frac{n \pi m^2 }{qE}}, \\
w_f &=& \frac{(qE)(qB)}{(2 \pi)^2} \sum_{n = 1}^{\infty}
\frac{1}{n} \coth \Bigl(\frac{n \pi B}{E} \Bigr) e^{- \frac{n \pi
m^2 }{qE}}.
\end{eqnarray}
Schwinger \cite{schwinger} used the term pair-production
probability for $w_b$ or $w_f$, and Itzykson and Zuber
\cite{itzykson} the pair-production rate, whereas Nikishov
\cite{nikishov} used only the first term of the series to denote
the pair-production rate.

The mean number of created pairs can be derived in the following
way. Let $P_0$ denote the probability for the vacuum-to-vacuum
transition and $P_1$ denote the probability for one-pair
production, then for the boson case one has the probability
conservation
\begin{equation}
1 = P_0 ( 1 + P_1 + P_1^2 + P_1^3 + \cdots ),
\end{equation}
and the mean number of created pairs
\begin{equation}
{\cal N}^b = P_0 ( P_1 + 2 P_1^2 + 3 P_1^3 + \cdots ).
\end{equation}
Therefore, one finds the relations
\begin{eqnarray}
P_0 &=& \frac{1}{1 + {\cal N}^b} = P^{nb}, \\
P_1 &=& \frac{{\cal N}^b}{ 1 + {\cal N}^b} = P^{b}.
\end{eqnarray}
In the case of fermions, one has the probability conservation
\begin{equation}
1 = P_0 ( 1 + P_1)
\end{equation}
due to the Pauli blocking, and the mean number of created pairs
\begin{equation}
{\cal N}^f = P_0 P_1.
\end{equation}
Thus one obtains the relations
\begin{eqnarray}
P_0 &=& 1 - {\cal N}_f = P^{nf}, \\
P_1 &=& \frac{{\cal N}^f}{ 1 - {\cal N}^f}.
\end{eqnarray}

\subsection{Instanton Method}

The space-dependent gauge has an advantage of readily being
applicable to static inhomogeneous fields. An inhomogeneous
electric field, for instance along the $z$-direction, has the
time-dependent gauge
\begin{equation}
A_z = \int E(z) dt = E(z) t.
\end{equation}
So it is difficult to apply the canonical method in the
time-dependent gauge. Now, in the space-dependent Coulomb gauge,
the mode-decomposed field equation takes the form
\begin{equation}
\Bigl [- \partial_z^2 + Q(z) \Bigr] \phi (z) = 0.
\end{equation}
Hence the problem reduces to a one-dimensional tunneling problem
to which we may apply the idea of the canonical method. In
general, there are two asymptotic regions with $Q(z) < 0$ and a
potential barrier $Q(z) > 0$ in between asymptotic regions. Then
the no-tunneling probability for the vacuum persistence for bosons
becomes
\begin{equation}
P^b = \frac{1}{1 + e^{- 2 S}},
\end{equation}
and for fermions
\begin{equation}
P^f = 1 - e^{- 2S},
\end{equation}
where $S$ is the total sum of all order instanton contributions
\begin{equation}
S = \sum_{k = 0}^{\infty} S_{(2k)}.
\end{equation}
The 0-loop or classical instanton action is
\begin{equation}
2 S_{(0)} = \oint Q^{1/2} (z),
\end{equation}
and $2 S_{(2k)}$ is the $k$-loop correction to the instanton
action.

\section{Pair Production from Charged Black Holes}

We now turn to the pair production by charged black holes. As a
charged black hole we consider the Reissner-Nordstr\"{o}m black
hole with the metric
\begin{equation}
ds^2 = - g(r) dt^2 + \frac{dr^2}{g(r)} + r^2 d \Omega^2_2.
\end{equation}
where
\begin{equation}
g(r) = 1 - \frac{2M}{r} + \frac{Q^2}{r^2}.
\end{equation}
Here $M$ and $Q$ are the mass and charge of the black hole. The
charge $Q$ also generates an electric field in the radial
direction given by the Coulomb potential
\begin{equation}
A_0 = \frac{Q}{r},\quad A_{i} = 0.
\end{equation}
Then the Klein-Gordon equation in the black hole background takes
the form
\begin{equation}
\Bigl[ - \frac{1}{g(r)}\Bigl\{\frac{\partial}{\partial t} - i
\frac{qQ}{r} \Bigr\}^2 + \frac{1}{r^2} \frac{\partial}{\partial r}
\Bigl\{r^2 g(r) \frac{\partial}{\partial r} \Bigr\} - \frac{{\bf
L}^2}{r^2} - m^2 \Bigr] \Phi = 0.
\end{equation}
It is convenient to introduce the tortoise coordinate
\begin{eqnarray}
r^* = \int \frac{dr}{g(r)} = r + \frac{r_+^2}{(r_+ - r_-)} \ln (r
- r_+) - \frac{r_-^2}{(r_+ - r_-)} \ln (r - r_-),
\end{eqnarray}
where $r_{\pm} = M \pm \sqrt{M^2 - Q^2}$ are the outer and inner
horizons. The field can be expanded into the spherical harmonics
\begin{equation}
\Phi (t, {\bf x}) = e^{i \omega t} \frac{\Psi_l(r)}{r} Y_{lm}
(\theta, \varphi), \label{wave ex}
\end{equation}
where it is assumed ${\rm Re}(\omega) > 0$ in all cases and ${\rm
Im} (\omega) = 0$ for the bounded states of stable motions and
${\rm Im} (\omega) > 0$ for the unbounded states of unstable
motions. Then the radial equations for the mode-decomposed field
(\ref{wave ex}) take the form
\begin{equation}
\Bigl[ - \frac{d^2}{d r^{*2}} + V_l (r)\Bigr] \Psi_l (r) =
\epsilon \Psi_{l} (r), \label{rn ef eq}
\end{equation}
where
\begin{eqnarray}
V_l (r) = g(r) \Bigl(m^2 + \frac{l(l+1)}{r^2} + \frac{2M}{r^3} -
\frac{2Q^2}{r^4}\Bigr) - \Bigl(\omega - \frac{qQ}{r} \Bigr)^2 +
(\omega^2 - m^2), \label{rn ef pot}
\end{eqnarray}
and
\begin{eqnarray}
\epsilon = \omega^2 - m^2.
\end{eqnarray}
Note that Eq. (\ref{rn ef eq}) has the same form as the
Regge-Wheeler equation for the axial perturbations of black holes
\cite{regge,zerilli}.

\subsection{Minkowski Spacetime}

In the Minkowski spacetime with $M = 0$ and $Q = 0$ in $g(r)$, the
effective potential for the radial motion reduces to
\begin{equation}
V_l (r) = \frac{2 \omega q Q }{r} + \frac{l(l+1) - q^2Q^2}{r^2}.
\label{m ef pot}
\end{equation}
Note that the effective potential (\ref{m ef pot}) of the
Klein-Gorodn equation is the same as that of the non-relativistic
system with a charge $2 \omega Q$ and effective angular momentum
$l'(l'+1) = l(l+1) - q^2 Q^2$, i.e.,
\begin{equation}
l' = - \frac{1}{2} + \sqrt{\Bigl(l + \frac{1}{2}\Bigr)^2 - q^2
Q^2}. \label{ef ang}
\end{equation}
The characteristics of states depend on the relative magnitude of
charge $qQ$ and angular momentum $l + 1/2$. The states for $\vert
qQ \vert < l + 1/2$ describe stable motions of bound states and
the hydrogen-like atoms with $Z < (l + 1/2)/ \alpha$ belong to
this class, where $Z$ is the nucleus charge and $\alpha = e^2$ the
fine structure constant \cite{bagrov,khalilov}. On the other hand,
for $qQ > l +1/2$, as in the case of heavy atoms with $Z
> (l+1/2)/\alpha$, the so-called 'centrifugal' potential (\ref{m ef pot})
becomes attractive and the effective angular momentum (\ref{ef
ang}) takes a complex value. Moreover, the electric field is
sufficiently strong enough to produce pairs of charged particles.
There are certain quasi-stationary or meta-stable states
describing the pair production.

We consider the interesting case of pair production by the Coulomb
field when $\vert qQ \vert > l + \frac{1}{2}$. The centrifugal
potential becomes attractive rather than repulsive, and there may
exist quasi-stationary states with  complex frequencies $\omega$
\begin{equation}
{\rm Re} (\omega) > 0,\quad {\rm Im} (\omega) > 0.
\end{equation}
Then the wave function (\ref{wave ex}) decays in time and
describes an outgoing wave for created particles by the Coulomb
field. One may understand the quasi-stationary states by a simple
analog of the quantum system having a potential well and finite
potential barrier. In our case the real part of the effective
potential consists of the attractive inverse square potential and
a repulsive Coulomb barrier for $qQ > 0$. Therefore, the
quasi-stationary state has a certain probability to tunnel the
barrier and decays in time. Such radial wave function is found
\begin{equation}
\Psi_l (r) = N r^{l'+1} e^{i \sqrt{\epsilon}r}M(l'+ 1 + i
\lambda', 2l' +2, - 2 i \sqrt{\epsilon}r), \label{unst wave}
\end{equation}
where $M$ is the confluent hypergeometric function and
\begin{equation}
\lambda' = \frac{\omega qQ}{\sqrt{\epsilon}}, \quad l' = -
\frac{1}{2} + i \sqrt{q^2Q^2 - (l+ \frac{1}{2})^2}
\end{equation}
From the asymptotic form
\begin{equation}
\Psi_l (r) = N r^{l'+1} e^{i \sqrt{\epsilon}r}\Biggl[ \frac{e^{-i
\pi (l'+ 1 + i \lambda')}}{\Gamma(l'+ 1 -i \lambda')} (- 2 i
\sqrt{\epsilon}r)^{- l' - 1 - i \lambda'} + \frac{1}{\Gamma(l'+ 1+
i \lambda')} (- 2 i \sqrt{\epsilon}r)^{- l' - 1 + i \lambda'} e^{-
2 i \sqrt{\epsilon}r} \Biggr], \label{asym}
\end{equation}
we see that, assuming ${\rm Re}(\sqrt{\epsilon}) > 0$, the first
term corresponds to an incoming wave and the second term to an
outgoing wave. We now require only outgoing waves and this
condition is prescribed by
\begin{equation}
l' + 1 - i \lambda' = - n_r,
\end{equation}
which makes first term vanish due to $\Gamma (- n_r) = \pm
\infty$. Then we obtain the complex energy eigenvalues
\begin{equation}
\omega = m \Biggl[1 + \Biggl(\frac{qQ}{n - (l +\frac{1}{2}) + i
\sqrt{q^2 Q^2 -(l+ \frac{1}{2})^2}} \Biggr)^2 \Biggr]^{-1/2}.
\label{com eig}
\end{equation}

We make a few comments. The consistency of quasi-stationary states
requires that ${\rm Re} (\omega)
> 0, {\rm Im} (\omega) > 0$ and ${\rm Re}(\sqrt{\epsilon}) > 0$,
which can be shown by a straightforward algebra. The result does
not depend on the relative sign of charges, and Eq. (\ref{com
eig}) holds true also for the opposite charges, $qQ < 0$. The
complex energy eigenvalues (\ref{com eig}) are the analytical
continuation of the real energy eigenvalues of bound states when
$qQ > l + \frac{1}{2}$. Moreover, the complex energy eigenvalues
are quantized and indeed describe the quasi-stationary or
meta-stable states that decay into infinity. More importantly, the
pair production rate given by the imaginary part of frequency is
proportional to $m \sqrt{qQ}$. This implies that the stronger the
electric field is, the more the pairs of charged particles are
produced.

\subsection{Charged Black Holes}

Now the radial equation for each mode is described by Eq. (\ref{rn
ef eq}). At sufficiently away from the black hole $(r/M \gg 1)$,
$r^* \approx r$ and Eq. (\ref{rn ef pot}) becomes approximately
\begin{equation}
V_{l} (r) \approx  \frac{2(\omega qQ - M)}{r} + \frac{l(l+1)-
q^2Q^2}{r^2}.
\end{equation}
Hence, Eq. (\ref{rn ef eq}) is approximately the same as that of
the Minkowski spacetime with the replacement of $\omega$ by
$\omega qQ - M$. Therefore, we are able to obtain the wave
function valid in the asymptotic regions
\begin{equation}
\Psi_l (r) = N r^{l'+1} e^{i \sqrt{\epsilon}r}M(l'+ 1 + i
\lambda_b, 2l' +2, - 2 i \sqrt{\epsilon}r), \label{rn unst wave}
\end{equation}
where
\begin{equation}
\lambda_b = \frac{\omega qQ - M}{\sqrt{\epsilon}}, \quad l' = -
\frac{1}{2} + i \sqrt{q^2Q^2 - (l+ \frac{1}{2})^2}.
\end{equation}
The outgoing wave is found by imposing the condition
\begin{equation}
l' + 1 - i \lambda' = - n_r,
\end{equation}
which removes the incoming wave. Finally, after some algebra, we
obtain the complex frequencies
\begin{equation}
\omega = m \Biggl[\frac{1 + \Bigl(\frac{qQ}{F} \Bigr)^2}{1 -
M^2F^2 \Bigl\{ 1 + \Bigl(\frac{qQ}{F} \Bigr)^2 \Bigr\}}
\Biggr]^{-1/2} + \frac{MqQ}{F^2\Bigl\{ 1 + \Bigl(\frac{qQ}{F}
\Bigr)^2 \Bigr\} }, \label{rn com eig}
\end{equation}
where
\begin{equation}
F = n - \Bigl(1 + \frac{1}{2} \Bigr) + i \sqrt{q^2Q^2 - \Bigl(l +
\frac{1}{2} \Bigr)^2}.
\end{equation}
In the limit of $M = 0$, we recover the result (\ref{com eig}) of
the Minkowski spacetime.

We make a few comments. There is a formal similarity of Eq.
(\ref{rn ef eq}) and the Regge-Wheeler equation for the axial
perturbations of the black hole \cite{regge,zerilli}. The boundary
condition for the purely outgoing wave leads to a discrete
spectrum of complex eigenfrequencies for both the quasi-stationary
states and quasi-normal modes
\cite{vishveshwara,press,chandrasekhar,bak}. However, the
quasi-stationary states exist even in the Minkowski spacetime,
whereas the quasi-normal modes vanish in the Minkowski spacetime.
This is ascribed to the fact that the pair production is a
universal phenomenon occurring in all spacetimes whereas the
quasi-normal modes are a consequence of the spacetime curvature.
More importantly, the decay rate, given by the imaginary part of
the frequency, has an additional contribution in proportion to the
mass of the hole. This implies that a charged black hole is a more
efficient mechanism for Schwinger pair production.

\subsection{Pair Production}

Finally, we apply the canonical method in Secs. III and IV to
calculate the pair-production rate from charged back holes. At the
spatial infinity $r \rightarrow + \infty$, $(r^* \rightarrow +
\infty)$, the effective potential vanishes, $V_l (r^* = \infty) =
0$, and at the outer horizon $r \rightarrow r_+$, $(r^*
\rightarrow - \infty)$, it has
\begin{equation}
V_l (r_+) = \omega^2 - m^2 - \Bigl(\omega - \frac{qQ}{r_+}
\Bigr)^2.
\end{equation}
Thus there is a potential barrier between $r^* = - \infty$ and
$r^* = + \infty$. We find the instanton action
\begin{eqnarray}
S_{l} (\omega; M, Q, m, q) = \int_{r_0^*}^{r_1^*} dr^* \sqrt{V_l
(r^*) - \epsilon} = \int_{r_0^*}^{r_1^*} dr  \frac{\sqrt{V_l (r^*)
- \epsilon}}{1 - \frac{2M}{r} + \frac{Q^2}{r^2}},
\end{eqnarray}
where $r^*_0$ and $r^*_1$ are the turning points of $V_l (r^*)$.
Then the mean number of produced boson pairs is
\begin{equation}
{\cal N}_{l} = e^{- 2 S_{l} (\omega)}. \label{bh mean}
\end{equation}
The mean number (\ref{bh mean}) is the leading term for the
pair-production rate and was derived also in Ref.
\cite{khriplovich}. The pair-production rate for bosons per volume
per time is given by
\begin{eqnarray}
2 {\rm Im} {\cal L}_e^b &=& (2 \sigma + 1) \sum_{l, \omega} (2l+1)
\ln \Bigl(1 + e^{ - 2 S_l (\omega)} \Bigr) \nonumber\\
&=& \frac{(2 \sigma + 1)}{2 \pi}  \sum_{l = 0, n = 1}^{\infty}
(-1)^{n+1} \frac{(2l+1)}{n}  \int d \omega  e^{ - 2n S_l
(\omega)},
\end{eqnarray}
and for fermions by
\begin{eqnarray}
2 {\rm Im} {\cal L}_e^f &=& - (2 \sigma + 1) \sum_{l, \omega}
(2l+1)
\ln \Bigl(1 - e^{ - 2 S_l (\omega)} \Bigr) \nonumber\\
&=& \frac{(2 \sigma + 1)}{2 \pi}  \sum_{l = 0, n = 1}^{\infty}
\frac{(2l+1)}{n}  \int d \omega  e^{ - 2n S_l (\omega)}.
\end{eqnarray}

\section{Conclusion}

We studied the pair production by inhomogeneous electromagnetic
fields, in particular, by charged black holes. We applied the
recently developed canonical method to find the pair-production
rate. For a static uniform electric field, the field equation can
be solved either in the time-dependent gauge or in the
space-dependent gauge. The concept of particle creation by an
external field applies directly to the field equation in the
time-dependent gauge. This time-dependent gauge, however, may not
be practically applicable for inhomogeneous fields due to the
mixed nature of potential. In the case of the space-dependent
gauge, the mode-decomposed field equations have tunneling
barriers, and the tunneling interpretation may be used for pair
production. In this interpretation the tunneling probability is
related with pair production and the no-tunneling probability with
the vacuum persistence, i.e., the vacuum-to-vacuum transition.
This canonical method can readily be applied to even inhomogeneous
fields. Further, the instanton action including all order
corrections leads to an accurate formula for the pair-production
rate.

Finally, we studied pair production by charged black holes. The
Klein-Gordon equation in a charged black hole background has the
form of the Regge-Wheeler equation. For a strong electric field,
the mode-decomposed field equations have the quasi-stationary
states that describe the purely outgoing waves for produced pairs.
Their complex frequencies determine the decaying rate for the
quasi-stationary states. We also noted a remarkable similarity
between the quasi-stationary states of created charged particles
and the quasi-normal modes of the axial or polar perturbations of
a black hole. The same boundary condition for purely outgoing
waves is used in both cases and, as a consequence, both wave
functions have a discrete spectrum of complex frequencies. We then
applied the canonical method to find the pair-production rate in
terms of the instanton action by charged black holes.

\acknowledgments We would like to thank R. Ruffini for many
stimulating discussions and also appreciate the warm hospitality
of ICRA, Pescara, Italy. The work of S.P. Kim was supported by the
Korea Science and Engineering Foundation under Grant No.
1999-2-112-003-5 and the work of D.N. Page was supported by
Natural Sciences and Engineering Council of Canada.

\end{document}